\documentclass[conference]{IEEEtran}
\usepackage[margin=0.7in]{geometry}

\usepackage[T1]{fontenc}

\usepackage{cite}
\usepackage[pdftex]{graphicx}
\graphicspath{{Figure/}}
\usepackage[cmex10]{amsmath}
\usepackage{amsthm}
\usepackage{amssymb}
\usepackage{algorithmic}
\usepackage{array}
\usepackage{color}
\usepackage{epstopdf}
\usepackage{amsfonts}

\usepackage{pgfplots}
\pgfplotsset{compat=1.3}
\usepackage{tikz}							% Tikz
\usetikzlibrary{shapes}
\usetikzlibrary{spy}
\usetikzlibrary{circuits}
\usetikzlibrary{arrows}

\newcommand{\vect}[1]{\boldsymbol{\mathrm{#1}}}
\newcommand{\mat}[1]{\boldsymbol{\mathrm{#1}}}

\newcommand{\tr}{\mathrm{tr}}

\newcommand{\diag}{\text{diag}}

\usepackage{mathtools}

\newtheorem{theorem}{Theorem}

\newtheorem{corollary}{Corollary}

\begin{document}

\title{Channel Extrapolation in FDD Massive MIMO:\\
	Theoretical Analysis and Numerical Validation}

\author{\IEEEauthorblockN{François Rottenberg\IEEEauthorrefmark{1}, Rui Wang\IEEEauthorrefmark{2}, Jianzhong Zhang\IEEEauthorrefmark{2} and Andreas F. Molisch\IEEEauthorrefmark{1},
}
\IEEEauthorblockA{\IEEEauthorrefmark{1}University of Southern California, Los Angeles, CA, USA,
	}
\IEEEauthorblockA{\IEEEauthorrefmark{2}Samsung Research America, Richardson, TX, USA
	}
}

\maketitle

\begin{abstract}
Downlink channel estimation in massive MIMO systems is well known to generate a large overhead in frequency division duplex (FDD) mode as the amount of training generally scales with the number of transmit antennas. Using instead an extrapolation of the channel from the measured uplink estimates to the downlink frequency band completely removes this overhead. In this paper, we investigate the theoretical limits of channel extrapolation in frequency. We highlight the advantage of basing the extrapolation on high-resolution channel estimation. A lower bound (LB) on the mean squared error (MSE) of the extrapolated channel is derived. A simplified LB is also proposed, giving physical intuition on the SNR gain and extrapolation range that can be expected in practice. The validity of the simplified LB relies on the assumption that the paths are well separated. The SNR gain then linearly improves with the number of receive antennas while the extrapolation performance penalty quadratically scales with the ratio of the frequency and the training bandwidth. The theoretical LB is numerically evaluated using a 3GPP channel model and we show that the LB can be reached by practical high-resolution parameter extraction algorithms. Our results show that there are strong limitations on the extrapolation range than can be expected in SISO systems while much more promising results can be obtained in the multiple-antenna setting as the paths can be more easily separated in the delay-angle domain.

\end{abstract}


\section{Introduction}\label{section:Introduction}
%\linespread{0.85}
Knowledge of Channel state information (CSI) at the transmitter (CSIT) is a fundamental prerequisite for operation of massive multiple-input - multiple-output (MIMO) communications systems. A massive MIMO system has a much larger number of antennas at the base station than the number of user antennas, implying that channel estimation is much less costly in uplink than in the downlink \cite{bjornson2017massive}. In time division duplex (TDD) systems, the base station (BS) can efficiently perform downlink channel estimation from uplink pilot transmission from the user equipments (UEs), since channel reciprocity holds as long as uplink and downlink transmission occurs within a coherence time of the channel, and {\em within the same frequency band}. However, in an FDD scenario, reciprocity cannot be exploited as different bands, usually separated by more than a coherence bandwidth, are used in uplink and downlink. On the other hand, estimation of the channel by downlink pilot transmission and feedback might result in a large overhead. To solve this dilemma, channel extrapolation from the uplink to the downlink band might provide a viable alternative.

Channel extrapolation in frequency was explored in \cite{Molisch2011}, which suggested estimation of the multipath components (MPCs) via high-resolution parameter estimation; based on the MPCs extrapolation over large bandwidths can be achieved. However, the single-antenna case that was considered in this paper showed poor performance. Vasisht et al. introduce a DL channel prediction method, which exploits the channel reciprocity between uplink and downlink channels to eliminate the need of UE CSI feedback in FDD systems \cite{Vasisht:2016}. In \cite{Ugurlu2016}, the authors proposed to acquire CSI through uplink pilots in combination with a limited feedback from downlink pilots. More recently, \cite{arnold2019enabling} considered extrapolation from uplink pilots using machine learning algorithms. Channel extrapolation in frequency also presents formal similarities to extrapolation in time. In contrast to frequency-domain extrapolation, channel prediction in time has been extensively investigated in the literature. A comprehensive review can be found in \cite{Hallen2007}. In \cite{Svantesson2006}, the authors proposed performance bounds for prediction in time of MIMO channels. They later extended their study to MIMO-OFDM channel estimation with interpolation and extrapolation being done both in time and frequency \cite{Larsen2009}. The authors make the observation that MIMO provides much longer prediction lengths than for SISO systems.

To provide understanding of the promise of low-overhead FDD massive MIMO systems, this paper investigates  the theoretical performance limits of channel extrapolation in frequency. First, we highlight the advantages of high-resolution channel estimation with respect to conventional least squares channel estimation resulting in SNR gain and extrapolation factor. Secondly, we formulate a theoretical LB on the MSE of the extrapolated channel, using a similar methodology as in \cite{Larsen2009}. The proposed LB differs from \cite{Larsen2009} as it takes into account elevation angles and the influence of the training pulse. Furthermore, a simplified LB is also proposed, giving more physical intuition on the extrapolation range and the SNR gain that can be expected in practice. The validity of the LB relies on the strong assumption that the paths are ``well separated". Thirdly, we analyze the performance of the theoretical LB by numerical simulations using a 3GPP channel model and showing that the LB can be reached by practical high-resolution parameter extraction algorithms. Our results show the very limited extrapolation range that can be achieved with SISO systems, while much more promising results are obtained in the MIMO setting as the paths can be more easily separated in the delay-angle domain.

\subsection{Notations}
Vectors and matrices are denoted by bold lowercase and uppercase letters, respectively. Superscripts $^*$, $^T$ and $^H$ stand for conjugate, transpose and Hermitian transpose operators. The symbols $\jmath$, $\tr$, $\mathbb{E}$, $\delta_n$, $\Im$ and $\Re$ denote the imaginary unit, trace, expectation, Kronecker delta, imaginary and real parts, respectively. The operator $\diag(\vect{a})$ returns a diagonal matrix with entries of vector $\vect{a}$ on its diagonal.

\section{System Model}
\label{section:system_model}

We consider FDD massive MIMO scenarios where each user has a single-antenna and transmits an uplink orthogonal training sequence. Thus, the estimation for different users becomes independent, and in particular the problem of extrapolating in frequency a SIMO channel. Moreover, to highlight the antenna arrays, the SISO case is studied in parallel.

Let us consider the transmission of a baseband equivalent pulse $s(t)$ \textit{a priori} known by the receiver. For the sake of concreteness, we consider in the following $s(t)$ being a root raised cosine (RRC) pulse with cutoff frequency is $\frac{1+\beta}{2T}$ where $0 \leq \beta \leq 1$ is the roll-off factor. Extension to an arbitrary training signal $s(t)$ is straightforward. We assume that the channel is quasi-static, \textit{i.e.}, constant for the duration of the transmission. $M$ denotes the number of antennas of the receive array. We assume that the propagation channel is composed of $L$ specular paths, where each path is completely characterized by its deterministic parameters: complex gain $\alpha_l=\alpha_l^R+\jmath \alpha_l^I$, delay $\tau_l$, azimuth angle $\phi_l$ and elevation angle $\theta_l$. Assuming that the ratio of the dimension of the array to the speed of light $c$ is much smaller than the inverse of the bandwidth of the signal, the complex baseband-equivalent of the received signal at antenna $m$ can be expressed as
\begin{align}
r_m(t)&=\sum_{l=1}^{L} \alpha_l {a}_m(\phi_l,\theta_l) s(t-\tau_{l}) +w_m(t), \label{eq:received_signal_TD}
\end{align}
where $w_m(t)$ is complex circularly symmetric white Gaussian noise and ${a}_m(\phi_l,\theta_l)$ is the pattern of antenna $m$ evaluated in the direction $(\phi_l,\theta_l)$. At the receiver, the signal at each antenna is pre-filtered and sampled at rate $1/T_s=K/T$ where the integer $K$ is the oversampling factor. The pre-filter has a unit frequency response in the bandwidth occupied by the signal of interest, \textit{i.e.}, $-\frac{1+\beta}{2T}\leq f\leq \frac{1+\beta}{2T}$, and is designed so that the noise is still white after filtering and sampling, \textit{i.e.}, $\mathbb{E}\left({w}_m[n]{w}^*_{m'}[n']\right)=\sigma_w^2\delta_{m-m'}\delta_{n-n'}$. The oversampling factor $K$ satisfies the relation $K\geq 1+\beta$, so that the useful signal is not impacted by aliasing after sampling. Defining ${r}_m[n]\triangleq{r}_m(nT_s)$ and ${w}_m[n]\triangleq{w}_m(nT_s)$, we obtain
\begin{align}
{r}_m[n]&=\sum_{l=1}^{L} \alpha_l {a}_m(\phi_l,\theta_l)   s(nT_s-\tau_{l})+{w}_m[n], \label{eq:received_signal_sampled_TD}
\end{align}
for $n=0,\hdots,N-1$ with $N$ being the number of observation samples. The SISO case can be seen as a special case of the SIMO case where each ray is completely characterized by its complex gain $\alpha_l$ and its delay $\tau_l$ while the information on the angles of arrival is lost. In the following, the index ``$m$" will be omitted when the single-antenna case is considered.

 \section{Channel estimation}
\label{section:channel_extrapolation}

%\begin{figure}[!t]
%	\centering
%	\resizebox{0.3\textwidth}{!}{%
%		\Large
%		\input{RRC_filter}
%	}
%	\vspace{-1em}
%	\caption{Frequency response of the transmit pulse.}
%	\label{fig:RRC_filter}
%\end{figure}

This section first describes the conventional low-resolution channel estimation technique used in most communication systems. In the light of its limitations, we detail our motivations for going towards high-resolution channel estimation. We define the channel frequency response evaluated at frequency $f$ and antenna $m$ as
\begin{align}
H_m(f)\triangleq\sum_{l=1}^{L} \alpha_l {a}_m(\phi_l,\theta_l) e^{-\jmath 2\pi f \tau_l} \label{eq:channel_frequency_response}.
\end{align}

\subsection{Conventional low-resolution estimation}

Taking the discrete-time Fourier transform of the received signal, we can rewrite (\ref{eq:received_signal_sampled_TD}) in the frequency domain as
\begin{align*}
R_m(f)&= H_m(f) S(f) +W_m(f),
\end{align*}
where $S(f)$, $R_m(f)$ and $W_m(f)$ are the discrete-time Fourier transforms of $r_m[n]$, $s[n]$ and $w_m[n]$. Conventional low-resolution algorithms such as least squares (LS) estimators perform a simple per-antenna estimation
\begin{align*}
\hat{H}^{\mathrm{LS}}_m(f)&=\frac{R_m(f)}{S(f)}=H_m(f)+\frac{W_m(f)}{S(f)}.
\end{align*}
We can easily see that the LS estimator is unbiased and is only limited by additive noise. Since the noise samples ${w}_m[n]$ are white and have variance $\sigma_w^2$, we can write
\begin{align*}
\text{MSE}_{\mathrm{LS}}(f)&\triangleq \mathbb{E} |\hat{H}^{\mathrm{LS}}_m(f)-{H}_m(f)|^2=\sigma_w^2\frac{N}{|S(f)|^2}.
\end{align*}
Since the transmit signal is a RRC pulse, we can explicitly write the expression of $S(f)$. Defining $\vect{s}\triangleq(s(0),...,s((N-1)T_s))^T$ and $\mathrm{SNR}\triangleq\frac{\|\vect{s}\|^2}{\sigma_w^2}$, we have% (see Fig.~\ref{fig:RRC_filter})
\begin{align*}
|S(f)|^2&=\begin{cases}
0 & \text{if } |f| \geq \frac{1+\beta}{2T}\\
\|\vect{s}\|^2 K & \text{if } |f| \leq \frac{1-\beta}{2T}
\end{cases}\\
\text{MSE}_{\mathrm{LS}}(f)&=\begin{cases}
\infty & \text{if } |f| \geq \frac{1+\beta}{2T}\\
\frac{1}{\mathrm{SNR}} \frac{N}{K} & \text{if } |f| \leq \frac{1-\beta}{2T}
\end{cases},
\end{align*}
where we omitted the transition band $\frac{1-\beta}{2T}\leq|f| \leq \frac{1+\beta}{2T} $ for the sake of simplicity. We can see that the MSE is infinite out of the bandwidth of the transmit signal, meaning that no extrapolation is possible. In practice, the MSE might not be infinite if simple linear extrapolation methods are used. However, most of these methods would have a very limited extrapolation range of the order of the coherence bandwidth of the channel, which can be related to the inverse of the maximal delay spread. For a delay spread of 2$\mu$s, this would correspond to only 500kHz extrapolation range.

In the signal band, the MSE of the LS estimator linearly degrades as a function of the ratio $\tilde{N}\triangleq \frac{N}{K}=\frac{NT_s}{T}$. We can see that $\tilde{N}$ actually corresponds to the number of time periods $T$ on which the received signal is being observed, including the delay spread of the channel and the transmit pulse duration. This can be intuitively seen as the number of delay coefficients that the LS estimator is trying to estimate. Indeed, because of its low resolution in time (of the order of $T$), the LS estimator does not resolve the specular paths but estimates instead its convolution with the transmit pulse, as shown in Fig.~\ref{fig:illustration_low_L}. To maximize the estimation performance, $\tilde{N}$ should be well chosen: big enough so that the observation window captures the useful training signal while being low enough just to capture useful signal and not noise. The gain obtained by limiting the number of coefficients $\tilde{N}$ is similar to an OFDM system that would convert its pilot-based frequency channel estimates to the time domain, truncate the obtained impulse response to $\tilde{N}$ coefficients and finally convert it back to frequency \cite{RottenbergConf3}. One can note that applying a LMMSE filter on the frequency estimates as in \cite{Edfors1998} might improve the performance but would require the \textit{a priori} knowledge of channel second order statistics.

\subsection{High-resolution estimation}
\label{section:high_resolution_motivation}

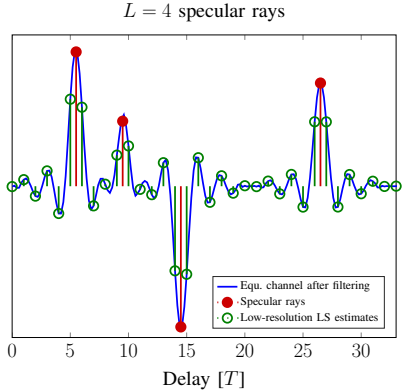
\begin{figure}[!t]
	\centering
	\resizebox{0.3\textwidth}{!}{%
		\LARGE
% Minimal pgfplots version: 1.3
%  http://www.mathworks.com/matlabcentral/fileexchange/22022-matlab2tikz
\begin{tikzpicture}

\begin{axis}[%
width=4.520833in,
height=3.565625in,
at={(0.758333in,0.48125in)},
scale only axis,
xmin=0,
xmax=33,
xlabel={Delay [$T$]},
ymin=-0.3,
ymax=0.3,
ytick={\empty},
title={$L=4$ specular rays},
legend style={at={(0.97,0.03)},anchor=south east,legend cell align=left,align=left,draw=white!15!black}
]
\addplot [color=blue,solid,line width=1.5pt]
  table[row sep=crcr]{%
0	-0.000737247146185985\\
0.2	0.00120101348818026\\
0.4	-0.00357406965785775\\
0.6	-0.00300184302394782\\
0.8	0.00731353509488781\\
1	0.0130448130073627\\
1.2	0.0154510676056086\\
1.4	0.0104103056296214\\
1.6	0.000762877884197371\\
1.8	-0.0111286727743205\\
2	-0.0196575851424859\\
2.2	-0.0210692608254092\\
2.4	-0.013374375544498\\
2.6	0.00157720882965652\\
2.8	0.0183787186058439\\
3	0.030411650665108\\
3.2	0.03095890698666\\
3.4	0.0181423242321577\\
3.6	-0.00645235549403073\\
3.8	-0.0336694052385588\\
4	-0.0540987708899045\\
4.2	-0.0541766096832489\\
4.4	-0.0314717623859952\\
4.6	0.0200905723044378\\
4.8	0.0950495162863095\\
5	0.172554700304167\\
5.2	0.236329924820725\\
5.4	0.269633882336669\\
5.6	0.264897193898936\\
5.8	0.2235252181028\\
6	0.156653496894261\\
6.2	0.0813733528799554\\
6.4	0.0149995679843497\\
6.6	-0.0289073913285797\\
6.8	-0.0459307830217134\\
7	-0.03894220906471\\
7.2	-0.0190170529540035\\
7.4	0.00262025937094022\\
7.6	0.0147637044618191\\
7.8	0.0148812863052731\\
8	0.00402840748210123\\
8.2	-0.00775536088797528\\
8.4	-0.0132398220782359\\
8.6	-0.00246764521552601\\
8.8	0.022664493726348\\
9	0.061607114927194\\
9.2	0.098436686598805\\
9.4	0.132606348599894\\
9.6	0.141673520374326\\
9.8	0.118593166134892\\
10	0.0798869166956643\\
10.2	0.0354107463108433\\
10.4	-0.00344569225084372\\
10.6	-0.0188147710196872\\
10.8	-0.0136042526173464\\
11	-0.00622195664740128\\
11.2	0.00625653358434178\\
11.4	0.0104753619253897\\
11.6	0.00750635287110317\\
11.8	-0.00461830894051627\\
12	-0.0167953770872146\\
12.2	-0.0238819439001073\\
12.4	-0.0178906237797737\\
12.6	-9.44240117415758e-05\\
12.8	0.0255008145702942\\
13	0.0466462218660829\\
13.2	0.052124008557596\\
13.4	0.0313956319955487\\
13.6	-0.0173996805806415\\
13.8	-0.08852778206918\\
14	-0.167732485674292\\
14.2	-0.236365743167168\\
14.4	-0.278556013002978\\
14.6	-0.279021405381179\\
14.8	-0.239136523832332\\
15	-0.174596324455093\\
15.2	-0.0955541866141366\\
15.4	-0.0228469148403468\\
15.6	0.031403696878464\\
15.8	0.0571530760804312\\
16	0.0565136191665449\\
16.2	0.0353593315607386\\
16.4	0.00672420709536558\\
16.6	-0.0190001365609398\\
16.8	-0.0324276592645\\
17	-0.0318877232181543\\
17.2	-0.0192223567933974\\
17.4	-0.00169551558228425\\
17.6	0.0140638088985101\\
17.8	0.022017950259607\\
18	0.0206607503937954\\
18.2	0.0115893884422892\\
18.4	-0.000721831035037277\\
18.6	-0.0109920025109379\\
18.8	-0.0160970442350131\\
19	-0.0137681783107184\\
19.2	-0.00755314613936601\\
19.4	0.00302465866428147\\
19.6	0.00387820500488871\\
19.8	-0.00140747099908298\\
20	0.000941215441819795\\
20.2	-0.00076289773572975\\
20.4	0.000685910537254133\\
20.6	-0.000665780859393201\\
20.8	0.000697366972308927\\
21	-0.000809179356345443\\
21.2	0.0011358995015827\\
21.4	-0.00293492267027516\\
21.6	-0.00213731976557842\\
21.8	0.00546729564444737\\
22	0.0101576726837621\\
22.2	0.0117484416064105\\
22.4	0.00810684716594503\\
22.6	0.00048913159569924\\
22.8	-0.0084643322615239\\
23	-0.0151760435073264\\
23.2	-0.0161239894552637\\
23.4	-0.0103259066408585\\
23.6	0.00125039496091139\\
23.8	0.0140941885345698\\
24	0.0233743318265148\\
24.2	0.0237916610625987\\
24.4	0.0139086609441582\\
24.6	-0.00489822339753828\\
24.8	-0.0259695184984361\\
25	-0.0413824928413986\\
25.2	-0.0419772251379776\\
25.4	-0.0229360727526732\\
25.6	0.0166348824384331\\
25.8	0.0702219887376762\\
26	0.127789685266284\\
26.2	0.175741733761388\\
26.4	0.203344118747131\\
26.6	0.203160081884686\\
26.8	0.175925954559761\\
27	0.127605095045292\\
27.2	0.0704071370069053\\
27.4	0.0164489826916913\\
27.6	-0.022749221511428\\
27.8	-0.0421652364304171\\
28	-0.041193102224498\\
28.2	-0.02616052090016\\
28.4	-0.00470536071456688\\
28.6	0.0137136701133072\\
28.8	0.0239890712341057\\
29	0.0231741830401199\\
29.2	0.0142974290805198\\
29.4	0.00104366850403099\\
29.6	-0.0101152502053025\\
29.8	-0.0163390811246418\\
30	-0.0149559357042593\\
30.2	-0.00869013156482751\\
30.4	0.000721417135783062\\
30.6	0.00786712774160115\\
30.8	0.0119967415524845\\
31	0.00989938185363535\\
31.2	0.00573734473730684\\
31.4	-0.00242138997592738\\
31.6	-0.00263386031899388\\
31.8	0.000813825459938848\\
32	-0.000460453300722117\\
32.2	0.000313705077933712\\
32.4	-0.000234280552079677\\
32.6	0.000184825412605509\\
32.8	-0.000151227154357433\\
33	0.000127005898400477\\
33.2	-0.000108779176720463\\
33.4	9.46123439880021e-05\\
33.6	-8.33202121083655e-05\\
33.8	7.41367905224027e-05\\
34	-6.65452949492595e-05\\
34.2	6.01845718545853e-05\\
34.4	-5.47946954332319e-05\\
34.6	5.01838710749202e-05\\
34.8	-4.62075149172885e-05\\
35	4.27545894340971e-05\\
};
\addlegendentry{{\normalsize Equ. channel after filtering}};

\addplot[ycomb,color=black!20!red,solid,line width=1.5pt,mark size=4.0pt,mark=*,mark options={solid,fill=black!20!red}] plot table[row sep=crcr] {%
5.5	0.266302166997231\\
9.5	0.128947348441411\\
14.5	-0.279021409111465\\
26.5	0.204588360570612\\
};
\addlegendentry{{\normalsize Specular rays}};

\addplot[ycomb,color=black!50!green,solid,line width=1.5pt,mark size=4.0pt,mark=o,mark options={solid}] plot table[row sep=crcr] {%
0	-0.000737247146185985\\
1	0.0130448130073627\\
2	-0.0196575851424859\\
3	0.030411650665108\\
4	-0.0540987708899045\\
5	0.172554700304167\\
6	0.156653496894261\\
7	-0.03894220906471\\
8	0.00402840748210123\\
9	0.061607114927194\\
10	0.0798869166956643\\
11	-0.00622195664740128\\
12	-0.0167953770872146\\
13	0.0466462218660829\\
14	-0.167732485674292\\
15	-0.174596324455093\\
16	0.0565136191665449\\
17	-0.0318877232181543\\
18	0.0206607503937954\\
19	-0.0137681783107184\\
20	0.000941215441819795\\
21	-0.000809179356345443\\
22	0.0101576726837621\\
23	-0.0151760435073264\\
24	0.0233743318265148\\
25	-0.0413824928413986\\
26	0.127789685266284\\
27	0.127605095045292\\
28	-0.041193102224498\\
29	0.0231741830401199\\
30	-0.0149559357042593\\
31	0.00989938185363535\\
32	-0.000460453300722117\\
33	0.000127005898400477\\
34	-6.65452949492595e-05\\
35	4.27545894340971e-05\\
};
\addlegendentry{{\normalsize Low-resolution LS estimates}};

\end{axis}
\end{tikzpicture}%
	}
	\vspace{-1em}
	\caption{For a scenario with few well separated rays, high-resolution channel estimation is preferable as it would have to estimate only $L=4$ coefficients instead of about $\tilde{N}\approx 30$ for a low-resolution LS estimator.}
	\label{fig:illustration_low_L}
\end{figure}

One could wonder if alternatives to conventional low-resolution channel estimation are possible. If, as depicted in Fig.~\ref{fig:illustration_low_L}, the channel only has a few well separated specular multipath components\footnote{The ``few well separated" assumption will be properly formalized in Section~\ref{section:Performance_limits}.}, \textit{i.e.}, $L<<\tilde{N}$, an intuitive reasoning suggests to go for high-resolution estimation of the $L$ different path parameters directly \cite{Molisch2011}. There are two main motivations for this: \textbf{SNR gain} and \textbf{extrapolation}. The \textbf{SNR gain} would come from two sources. First, the receiver only has to estimate $L$ complex coefficients rather than $\tilde{N}$, resulting in a potential SNR gain of $\frac{\tilde{N}}{L}$ with respect to LS estimation. One should note that this gain is stronger than the potential gain of using a frequency correlation filter as in \cite{Edfors1998}. It does not come from simply assuming that the channel impulse response has a finite length $\tilde{N}T$ as in \cite{RottenbergConf3} but from its sparsity. Secondly, the received signal at each antenna can be coherently combined to jointly estimate and separate all path parameters instead of performing per-antenna independent channel estimation as in the LS case, which results in a potential total SNR gain of a factor $\frac{M\tilde{N}}{L}$ with respect to LS estimation. One should here also note that improved channel estimators making using of correlation in the spatial domain through the antennas could also achieve a similar gain. By definition, low-resolution estimators are restricted to the bandwidth occupied by the training signal. However, high-resolution estimates of the path parameters allow for simple channel \textbf{extrapolation} in frequency possibly very far from the initial band of the training signal. If we denote by $\hat{\tau}_l$, $\hat{\phi}_l$, $\hat{\theta}_l$ and $\hat{\alpha}_l$ the high-resolution estimates of ${\tau}_l$, ${\phi}_l$, ${\theta}_l$ and ${\alpha}_l$ respectively, we can write the expression of the extrapolated channel as
\begin{align}
\hat{H}_m(f,\hat{\vect{\psi}})=\sum_{l=1}^{L} \hat{\alpha}_l {a}_m(\hat{\phi}_l,\hat{\theta}_l) e^{-\jmath 2\pi f \hat{\tau}_l} \label{eq:channel_frequency_extrapolated}.
\end{align}
Of course, intuitive reasoning tells us that the extrapolated channel will suffer from the estimation errors on the path parameters, especially as the extrapolation factor becomes large. We also assume here that the underlying assumptions of the model in (\ref{eq:received_signal_TD}) are still holding, \textit{i.e.}, the inverse of the extrapolation range is much smaller than the inverse of the dimension of the array to the speed of light, which implies that the antenna array response is the same between uplink and downlink. Furthermore, we assume that the parameters of the MPCs are independent of frequency. This is well fufilled in most practical situations, since the range over which these parameters change is on the order of $10\%$ of the carrier frequency, which is much larger than the extrapolation range we can usually obtain, see Section~\ref{section:Numerical_validation}.

\section{Performance Analysis}
\label{section:Performance_limits}

To theoretically formalize the two potential gains of high-resolution channel estimation, we will in a first step derive the Fisher information matrix of the estimated path parameters. The second step will consist in deriving a range on the MSE of the extrapolated channel frequency response. In a third step, a simplified LB will be proved giving much more physical intuition. Finally, the previous results will be particularized to the SISO case.

\subsection{Fisher Information Matrix}

Let us define the vector $\vect{r}\in \mathbb{C}^{NM\times 1}$ as containing all received samples for all antennas and observation samples. We also define the vector $\vect{\psi}\in \mathbb{R}^{5L\times 1}$ as containing the $5L$ real-valued path parameters. Given the independence of the noise samples, the log-likelihood of vector $\vect{r}$ becomes
\begin{align*}
L\left(\vect{r};\vect{\psi}\right)&=\sum_{n=0}^{N-1 }\sum_{m=1}^{M} L\left({r}_m[n];\vect{\psi}\right).
\end{align*}
The elements of the Fisher information matrix $\mat{I}_{\vect{\psi}} \in \mathbb{R}^{5L \times 5L}$ can be obtained from the log-likelihood function as \cite{kay1993fundamentals}
\begin{align}
\left[\mat{I}_{\vect{\psi}}\right]_{u,v}&=-\mathbb{E} \left( \frac{\partial^2 L\left(\vect{r};\vect{\psi}\right)}{\partial \psi_u \partial\psi_v} \right)\label{eq:I_psi_uv_raw}\\
&=-\sum_{n=0}^{N-1 }\sum_{m=1}^{M} \mathbb{E} \left( \frac{\partial^2 L\left({r}_m[n];\vect{\psi}\right)}{\partial \psi_u \partial\psi_v} \right)\nonumber,
\end{align}
where the expectation is taken over the noise distribution. Since the random variable $r_m[n];\vect{\psi}$ follows a circularly symmetric complex normal distribution with variance $\sigma_w^2$ and mean
%\begin{align*}
$\mu_{m,n}\triangleq\sum_{l=1}^{L}{{\alpha}_l}  {a}_m(\phi_l,\theta_l)   s(nT_s-\tau_{l})$,
%\end{align*}
equation (\ref{eq:I_psi_uv_raw}) can be rewritten as
\begin{align}
\left[\mat{I}_{\vect{\psi}}\right]_{u,v}&=\frac{2}{\sigma_w^2}\sum_{n=0}^{N-1 }\sum_{m=1}^{M} \Re \left\lbrace \frac{\partial \mu_{m,n}^*}{\partial \psi_u} \frac{\partial \mu_{m,n}}{\partial \psi_v} \right\rbrace. \label{eq:Fisher_info_matrix_element}
\end{align}
Separating the different path parameters in vector $\vect{\psi}$, we can partition the full $5L\times 5L$ Fisher information matrix in 25 submatrices, each of dimension $L\times L$, as
\begin{align}
\mat{I}_{\vect{\psi}}&=\frac{2}{\sigma_w^2}\begin{pmatrix}
\mat{I}_{\tau \tau}       & \mat{I}_{\tau \phi}       & \mat{I}_{\tau \theta}        & \mat{I}_{\tau \alpha^R}       & \mat{I}_{\tau \alpha^I}\\
\mat{I}_{\tau \phi}^T     & \mat{I}_{\phi \phi}       & \mat{I}_{\phi \theta}        & \mat{I}_{\phi \alpha^R}       & \mat{I}_{\phi \alpha^I}\\
\mat{I}_{\tau \theta}^T   & \mat{I}_{\phi \theta}^T   & \mat{I}_{\theta \theta}      & \mat{I}_{\theta \alpha^R}     & \mat{I}_{\theta \alpha^I}\\
\mat{I}_{\tau \alpha^R}^T & \mat{I}_{\phi \alpha^R}^T & \mat{I}_{\theta \alpha^R}^T  & \mat{I}_{\alpha^R \alpha^R}   & \mat{I}_{\alpha^R \alpha^I}\\
\mat{I}_{\tau \alpha^I}^T & \mat{I}_{\phi \alpha^I}^T & \mat{I}_{\theta \alpha^I}^T  & \mat{I}_{\alpha^R \alpha^I}^T & \mat{I}_{\alpha^I \alpha^I}\\
\end{pmatrix}. \label{eq:Fisher_information_matrix}
\end{align}
Defining $\dot{s}(t)\triangleq\frac{d s(t)}{dt}$, $\dot{a}_{m,\phi}(\phi,\theta)\triangleq\frac{d a_m(\phi,\theta)}{d \phi}$ and $\dot{a}_{m,\theta}(\phi,\theta)\triangleq\frac{d a_m(\phi,\theta)}{d \theta}$, we can write the partial derivatives appearing in (\ref{eq:Fisher_info_matrix_element}) as
\begin{align*}
\frac{d \mu_{m,n}}{d \tau_l}&= -\alpha_l{a}_m(\phi_l,\theta_l)   \dot{s}(nT_s-\tau_l) \\
\frac{d \mu_{m,n}}{d \phi_l}&= \alpha_l \dot{a}_{m,\phi}(\phi,\theta)  {s}(nT_s-\tau_l) \\
\frac{d \mu_{m,n}}{d \theta_l}&= \alpha_l \dot{a}_{m,\theta}(\phi,\theta)  {s}(nT_s-\tau_l) \\
\frac{d \mu_{m,n}}{d \alpha^R_l}&= {a}_m(\phi_l,\theta_l)   s(nT_s-\tau_l)\\
\frac{d \mu_{m,n}}{d \alpha^I_l}&= \jmath{a}_m(\phi_l,\theta_l)   s(nT_s-\tau_l).
\end{align*}
Inserting these partial derivatives in (\ref{eq:Fisher_info_matrix_element}) and for a specific array pattern $a_m(\phi,\theta)$, the Fisher information matrix in (\ref{eq:Fisher_information_matrix}) can be easily constructed. In the following, we will make the following assumption.

$\mathbf{(As1)}$: the Fisher information matrix $\mat{I}_{\vect{\psi}}$ is nonsingular.

In practice, a rank deficiency of $\mat{I}_{\vect{\psi}}$ could arise if two rays, or more, become extremely close in delay and angle, which would cause the determinant of $\mat{I}_{\vect{\psi}}$ to go to zero. A solution in this case can be to replace the two correlated rays with one ray whose amplitude is the sum of the amplitudes of the components. It is intuitive that this operation will not cause a large information loss if the rays are close enough.

\subsection{Lower bound on the MSE of the extrapolated channel}
Let us denote by $\hat{\vect{\psi}}\in \mathbb{R}^{5L\times 1}$ an unbiased estimator of $\vect{\psi}$ with covariance matrix
\begin{align*}
\mat{C}_{\hat{\vect{\psi}}}=\mathbb{E}\left(\left(\vect{\psi}-\hat{\vect{\psi}}\right)\left(\vect{\psi}-\hat{\vect{\psi}}\right)^T\right),
\end{align*}
where the expectation is taken over the noise distribution. The Cramer-Rao lower bound (CRLB) tells us that the matrix $\mat{C}_{\hat{\vect{\psi}}} -  \mat{I}_{\vect{\psi}}^{-1}$ is positive semidefinite, which implies that
%\begin{align}
$\vect{g}^H \mat{C}_{\hat{\vect{\psi}}} \vect{g} \geq \vect{g}^H\mat{I}_{\vect{\psi}}^{-1} \vect{g}$ % \label{eq:CRLB}
%\end{align}
for every vector $\vect{g}\in \mathbb{C}^{5L\times 1}$. If vector $\vect{g}$ is chosen as an all zero vector except a one at $u$-th entry, we get a LB for the variance of the estimated parameter $\psi_u$. The CRLB only provides a LB on the variance of the estimated parameters while we are interested on the variance on the error of the extrapolated channel, which we define as
\begin{align*}
\text{MSE}_m(f,\hat{\vect{\psi}})\triangleq{\mathbb{E} \left|\hat{H}_m(f,\hat{\vect{\psi}})-{H}_m(f)\right|^2}.
\end{align*}
%where $\vect{\alpha}=(\alpha_1,...,\alpha_L)$.
To obtain a performance limit, we would like to lower bound the MSE by a certain quantity $\mathrm{LB}_m(f,\hat{\vect{\psi}})$ so that
\begin{align*}
\text{MSE}_m(f,\hat{\vect{\psi}})\geq \mathrm{LB}_m(f,\vect{\psi}),
\end{align*}
where $\mathrm{LB}_m(f,\vect{\psi})$ would only depend on deterministic parameters. The following theorem gives a closed-form expression of the LB on the extrapolated channel as a function of the path parameters $\vect{\psi}$ and the extrapolated frequency $f$.

\begin{theorem} \label{theorem:LB}
	Under $\mathbf{(As1)}$, the LB on the MSE of the extrapolation error $\mathrm{LB}_m(f,\vect{\psi})$ for any unbiased estimator $\hat{H}_m(f,{\vect{\psi}})$ can be expressed as
	\begin{align*}
	{\mathrm{LB}}_m(f,\vect{\psi})&\triangleq\vect{g}^H_{m,f,\vect{\psi}}\mat{I}_{\vect{\psi}}^{-1}\vect{g}_{m,f,\vect{\psi}},
	\end{align*}
	where we defined the vectors
	\begin{align*}
	\vect{g}_{m,f,\vect{\psi}}^T&\triangleq\begin{pmatrix}
	\vect{a}^T_{m,f,\tau} & \vect{a}_{m,f,\phi}^T & \vect{a}_{m,f,\theta}^T & \vect{a}_{m,f,\alpha^R}^T & \vect{a}_{m,f,\alpha^I}^T
	\end{pmatrix}\\
	\vect{a}_{m,f,\tau}&\triangleq-\jmath 2\pi f\mat{D}_{\alpha}\mat{D}_{\tau}\begin{pmatrix}
	{a}_m({\phi}_1,{\theta}_1)& \hdots & {a}_m({\phi}_L,{\theta}_L)
	\end{pmatrix}^T\\
	\vect{a}_{m,f,\phi}&\triangleq\mat{D}_{\alpha}\mat{D}_{\tau}\begin{pmatrix}
	\dot{a}_{m,\phi}({\phi}_1,{\theta}_1)& \hdots & \dot{a}_{m,\phi}({\phi}_L,{\theta}_L)
	\end{pmatrix}^T\\
	\vect{a}_{m,f,\theta}&\triangleq\mat{D}_{\alpha}\mat{D}_{\tau}\begin{pmatrix}
	\dot{a}_{m,\phi}({\theta}_1,{\theta}_1)& \hdots & \dot{a}_{m,\theta}({\phi}_L,{\theta}_L)
	\end{pmatrix}^T\\
	\vect{a}_{m,f,\alpha^R}&\triangleq\mat{D}_{\tau}\begin{pmatrix}
	{a}_m({\phi}_1,{\theta}_1)& \hdots & {a}_m({\phi}_L,{\theta}_L)
	\end{pmatrix}^T\\
	\vect{a}_{m,f,\alpha^I}&\triangleq \jmath \mat{D}_{\tau} \begin{pmatrix}
	{a}_m({\phi}_1,{\theta}_1)& \hdots & {a}_m({\phi}_L,{\theta}_L)
	\end{pmatrix}^T,
	\end{align*}
	with $\mat{D}_{\tau}\triangleq\diag(e^{-\jmath 2\pi f \tau_{1}},...,e^{-\jmath 2\pi f \tau_{L}})$ and $\mat{D}_{\alpha}\triangleq\diag(\alpha_1,...,\alpha_L)$.
	
	\begin{proof}
		Proofs are omitted in this document due to space constraints, we refer to the journal paper for complete proofs.
	\end{proof}
\end{theorem}

\subsection{Separated rays}

The LB of Theorem~\ref{theorem:LB} is interesting and is in closed-form, which allows to easily evaluate it numerically. However, it requires the inversion of the Fisher information matrix and does not provide much intuition on the exact SNR gain and extrapolation factor that we can expect. To further characterize and try to get more insight on ${\mathrm{LB}}_m(f,\vect{\psi})$, let us first define the following vectors in order to introduce assumptions $\mathbf{(As2)-(As3)}$
\begin{align*}
\vect{s}_{l}&\triangleq\begin{pmatrix}
s(-\tau_l)& \hdots & s((N-1)T_s-\tau_l)
\end{pmatrix}^T\in \mathbb{C}^{N\times 1} \\
\dot{\vect{s}}_{l}&\triangleq\begin{pmatrix}
\dot{s}(-\tau_l)& \hdots & \dot{s}((N-1)T_s-\tau_l)
\end{pmatrix}^T\in \mathbb{C}^{N\times 1}\\
\vect{a}_l&\triangleq\begin{pmatrix}
{a}_1(\phi_{l},\theta_{l})& \hdots &{a}_M(\phi_{l},\theta_{l})
\end{pmatrix}^T\in \mathbb{C}^{M\times 1}\\
\dot{\vect{a}}_{l,\phi}&\triangleq\begin{pmatrix}
\dot{a}_{1,\phi}(\phi_l,\theta_l)& \hdots &\dot{a}_{M,\phi}(\phi_l,\theta_l)
\end{pmatrix}^T\in \mathbb{C}^{M\times 1}\\
\dot{\vect{a}}_{l,\theta}&\triangleq\begin{pmatrix}
\dot{a}_{1,\theta}(\phi_l,\theta_l)& \hdots &\dot{a}_{M,\theta}(\phi_l,\theta_l)
\end{pmatrix}^T\in \mathbb{C}^{M\times 1}.
\end{align*}

$\mathbf{(As2)}$: separation of the $L$ specular rays in delay, azimuth angle and/or elevation angle. We assume that, for each pair of rays $l,l'$ ($l\neq l'$), at least one of the following two relationships is verified:

(1) Separation in delay:
\begin{align}
\vect{s}_{l}^H\vect{s}_{l'}=\dot{\vect{s}}_{l}^H\dot{\vect{s}}_{l'}=\dot{\vect{s}}_{l}^H{\vect{s}}_{l'}=0. \label{eq:separation_in_delay}
\end{align}

(2) Separation in azimuth and/or elevation angle:
\begin{align*}
\vect{a}_l^H\vect{a}_{l'}=\dot{\vect{a}}_{l,\theta}^H\dot{\vect{a}}_{l,\theta}=\dot{\vect{a}}_{l,\phi}^H\dot{\vect{a}}_{l,\phi} =\dot{\vect{a}}_{l,\theta}^H\dot{\vect{a}}_{l}=\dot{\vect{a}}_{l,\phi}^H\dot{\vect{a}}_{l}=\dot{\vect{a}}_{l,\phi}^H \dot{\vect{a}}_{l',\theta} =0.
\end{align*}

The assumption $\mathbf{(As2)}$ is a strong assumption, whose accuracy will typically depend on different parameters. The specular paths will generally become more separated in delay as the bandwidth of $s(t)$ increases, inducing higher resolution in time. Similarly, the separation in azimuth and elevation will be improved as the number of antenna elements $M$ is increased. More generally, the validity of $\mathbf{(As2)}$ will depend on the pulse shape $s(t)$ and the array pattern $a_m(\phi,\theta)$.

$\mathbf{(As3)}$: the array pattern has even or odd symmetry in space implying that
\begin{align*}
\dot{\vect{a}}_{l,\phi}^H {\vect{a}}_{l}=\dot{\vect{a}}_{l,\theta}^H {\vect{a}}_{l}=0.
\end{align*}

The following corollary gives a particularization of the LB of Theorem~\ref{theorem:LB} under additional assumptions $\mathbf{(As2)}-\mathbf{(As3)}$ and for the MSE averaged over the receive antennas, \textit{i.e.},
\begin{align*}
\text{MSE}(f,\hat{\vect{\psi}})\triangleq \frac{1}{M}\sum_{m=1}^M\text{MSE}_m(f,\hat{\vect{\psi}}).
\end{align*}

\begin{corollary} \label{corollary:well_separated_rays}
	Under $\mathbf{(As2)}-\mathbf{(As3)}$, the expression of the LB of Theorem~\ref{theorem:LB} averaged over the receive antennas simplifies to
	\begin{align*}
	{\mathrm{LB}}(f,\vect{\psi})&\triangleq\frac{1}{M}\sum_{m=1}^{M}{\mathrm{LB}}_m(f,\vect{\psi})\\
	&=\frac{1}{\mathrm{SNR}}\underbrace{\frac{L}{M}}_{\mathrm{SNR\ gain}}(\underbrace{2}_{\mathrm{Loss\ factor}}+\underbrace{\frac{1}{2}\left(\frac{f}{\sigma_F}\right)^2}_{\mathrm{Extrapolation\ factor}}),
	\end{align*}
	where $\sigma_F^2$ is the mean squared bandwidth of the transmit signal
	\begin{align*}
	\sigma_F^2&\triangleq\frac{\|\dot{\vect{s}}\|^2}{(2\pi)^2\|\vect{s}\|^2}=\frac{\int_{f}f^2|S(f)|^2df}{\int_{f}|S(f)|^2df}.
	%=\frac{\int_{f}^{1/(2T_s)}f^2|S(f)|^2df}{\int_{-1/(2T_s)}^{1/(2T_s)}|S(f)|^2df}.
	\end{align*}
%	\begin{proof}
%		The proof is omitted due to space constraints.
%	\end{proof}
\end{corollary}

By adding some assumptions, the LB proposed in Theorem~\ref{theorem:LB} can be greatly simplified and Corollary~\ref{corollary:well_separated_rays} provides much insight into the physical meaning of the different terms of the LB. We can clearly identify the two main advantages of high-resolution channel estimation. As expected, a SNR gain of a factor $\frac{M\tilde{N}}{L}$ can be observed with respect to LS estimation. This gain comes from two contributions: the array gain $M$ and the fact that we estimate only $L$ channel paths instead of $\tilde{N}$ in the LS case. However, a loss factor of 2 appears, coming from the penalty of estimating the real and imaginary gains, the azimuth and the elevations angles of each path. Secondly, the channel can be extrapolated in frequency at the cost of a MSE penalty that quadratically scales with the ratio ${f}/{\sigma_F}$ where the denominator indicates that the extrapolation factor can be quantified in multiples of the signal bandwidth.

Furthermore, it is interesting to see that the dependence of ${\mathrm{LB}}(f,\vect{\psi})$ on the path parameters $\vect{\psi}$ vanish under $\mathbf{(As2)}-\mathbf{(As3)}$. This is in part explained by the fact that each path is well separated, which cancels the interdependence between the paths paramters in the expression of the LB. Additionally, the channel frequency response is evaluated in the direction of the incoming specular waves, canceling the dependence in the parameters of each path.

% ML not sensititive to the obeservation window

\subsection{Single-input-single-output}

The specialization of the above results to the SISO case is straightforward. As the angles of arrival are not resolved, the Fisher information matrix becomes a $3L\times 3L$ matrix. To simplify the LB, we introduce the following adaptation of $\mathbf{(As2)}$ to the SISO case:

$\mathbf{(As2')}$: separation of the $L$ specular rays in delay. We assume that, for each pair of rays $l,l'$ ($l\neq l'$), the condition (\ref{eq:separation_in_delay}) is verified.

\begin{corollary} \label{corollary:SISO_case}
	The LB on the channel extrapolation error for any unbiased estimator $\hat{H}(f,\hat{\vect{\psi}})$ in the SISO case is
	\begin{align*}
	{\mathrm{LB}}_{\mathrm{SISO}}(f,\vect{\psi})&\triangleq\vect{g}^H_{f,\vect{\psi}}\mat{I}_{\vect{\psi},\mathrm{SISO}}^{-1}\vect{g}_{f,\vect{\psi}},
	\end{align*}
	where 
	\begin{align*}
	\vect{g}_{f,\vect{\psi}}^T&\triangleq\begin{pmatrix}
	\vect{a}_{f,\tau}^T & \vect{a}_{f,\alpha^R}^T & \vect{a}_{f,\alpha^I}^T
	\end{pmatrix}\\
	\vect{a}_{f,\tau}&\triangleq-\jmath 2\pi f \mat{D}_{\alpha}\begin{pmatrix}
	e^{-\jmath 2\pi f \tau_{1}}& \hdots & e^{-\jmath 2\pi f \tau_{L}}
	\end{pmatrix}^T\\
	\vect{a}_{f,\alpha^R}&\triangleq\begin{pmatrix}
	e^{-\jmath 2\pi f \tau_{1}}& \hdots & e^{-\jmath 2\pi f \tau_{L}}
	\end{pmatrix}^T\\
	\vect{a}_{f,\alpha^I}&\triangleq\jmath\begin{pmatrix}
	e^{-\jmath 2\pi f \tau_{1}}& \hdots & e^{-\jmath 2\pi f \tau_{L}}
	\end{pmatrix}^T.
	\end{align*}
	Under $\mathbf{(As2')}$, the LB simplifies to
	\begin{align*}
	{\mathrm{LB}}_{\mathrm{SISO}}(f,\vect{\psi})=\underbrace{\frac{L}{\mathrm{SNR}}}_{\mathrm{SNR\ gain}}(\underbrace{1}_{\mathrm{Loss\ factor}}+\underbrace{\frac{1}{2}\left(\frac{f}{\sigma_F}\right)^2}_{\mathrm{Extrapolation\ factor}}).
	\end{align*}
%	\begin{proof}
%		We refer to the journal paper for the proof.
%	\end{proof}
\end{corollary}

As could be expected, the only SNR gain now comes from estimating $L$ coefficients rather than $\tilde{N}$. The extrapolation factor is the same as in the SIMO case. One can note that the loss factor is only one versus two in the SIMO case as the azimuth and elevation angles of each path are not estimated. The main difference of the SISO case with the SIMO case is the fact that far fewer observations of the channel are available, especially compared to a massive MIMO scenario with a large $M$. This not only eliminates the array gain but also makes $\mat{I}_{\vect{\psi},\mathrm{SISO}}$ more ill-conditioned as the rays can only be separated in the delay domain. As a result, $\mathbf{(As2')}$ is a stronger assumption than $\mathbf{(As2)}$ and might only be valid for a small number of rays $L$ and/or a very large bandwidth. These observations tend to strongly limit the potential gains of high-resolution channel estimation in SISO systems \cite{Molisch2011}.

\section{Numerical Validation}
\label{section:Numerical_validation}

This section aims at assessing the accuracy of the theoretical LB of the extrapolated channel through simulations. In the simulations, we used a RRC pulse shape $s(t)$ with roll-off factor $\beta=0.2$. The center frequency is set to 3.5 GHz. We consider a rectangular planar array of antennas at receive side with an inter-antenna element spacing of $\lambda_c/2$. The antenna elements have an isotropic pattern implying that $\mathbf{(As3)}$ holds. The SISO performance will also be considered according to the description in previous sections. Conventional LS estimation will be considered as a benchmark. For extracting the MPC, we use the SAGE algorithm presented in \cite{Fleury1999} straightforwardly extended to extract elevation angles. The performance of the algorithm was averaged over multiple noise realizations.
\begin{figure}[!t]
	\centering
	\resizebox{0.45\textwidth}{!}{%
		\Large
% Minimal pgfplots version: 1.3
%  http://www.mathworks.com/matlabcentral/fileexchange/22022-matlab2tikz
\definecolor{mycolor1}{rgb}{0.00000,0.44700,0.74100}%
\begin{tikzpicture}

\begin{axis}[%
width=4.520833in,
height=1.492587in,
at={(0.758333in,0.48125in)},
scale only axis,
xmin=-1,
xmax=3.14159265358979,
xlabel={Azimuth $\phi$ [rad]},
ymin=0,
ymax=3.14159265358979,
ylabel={Elevation $\theta$ [rad]}
]
\addplot [color=blue,line width=1.5pt,mark size=4.0pt,only marks,mark=x,mark options={solid},forget plot]
  table[row sep=crcr]{%
1.10714871779409	1.43742864902017\\
0.657528011581673	2.1978940911635\\
2.89379461442712	1.09381025837938\\
-0.811102681171197	1.76612097349285\\
1.79304614048497	0.32967989353401\\
3.06750030200244	0.482158992096567\\
-0.699537754063465	1.67537992124024\\
1.22246861155832	0.369070597206976\\
2.30102662595464	2.26605949566926\\
1.53253215462311	2.83318742815342\\
0.813007798495736	1.77439856107439\\
1.68485215382081	1.65285138025446\\
0.313847343702935	2.23634354427491\\
2.09169596533384	0.620304048268217\\
1.12407604257964	1.87139095699228\\
-0.808914965772551	1.61587991631172\\
2.05175836771008	1.30793901575541\\
1.05240280212522	0.598199712449126\\
-0.786164330984782	1.14129719128996\\
0.196591640295619	2.26813988397916\\
2.45256551317767	0.64846711881751\\
};
\end{axis}

\begin{axis}[%
width=4.520833in,
height=1.492587in,
at={(0.758333in,2.554288in)},
scale only axis,
xmin=0,
xmax=2e-06,
xlabel={Delay [s]},
ymin=0,
ymax=0.15,
ylabel={Gain [linear]}
]
\addplot[ycomb,color=mycolor1,solid,line width=1.5pt,mark size=4.0pt,mark=o,mark options={solid}] plot table[row sep=crcr] {%
0	9.999999999e-11\\
2.06865350890985e-07	0.0720925269986881\\
1.12031419321078e-06	0.0544632648528015\\
9.53304865868405e-07	0.0527548139016662\\
1.5293196812267e-08	0.0410007424296801\\
1.37097407069599e-08	0.027487188808278\\
2.73824829122872e-07	0.0320173195724815\\
7.31669280201506e-08	0.0711147380179798\\
4.05992029524148e-07	0.0274069301220075\\
1.02325585881178e-06	0.0584078103641195\\
4.63012591353619e-07	0.0721655361113138\\
4.76979928593672e-07	0.106408540653015\\
7.35622028218821e-07	0.0744621716136322\\
3.32767390449043e-07	0.0498322211242097\\
7.67946970540328e-07	0.0807427581354222\\
1.90091100807234e-06	0.0471641449584828\\
4.04788453784699e-08	0.0318676277120648\\
7.55281376597191e-08	0.0326637072974101\\
1.19204196854928e-06	0.0242556532807099\\
5.2032583225211e-07	0.0246081289759268\\
1.88594621609118e-07	0.0190841749701104\\
};
\end{axis}
\end{tikzpicture}%
	}
	\vspace{-1em}
	\caption{Generated set of parameters $(\alpha_l,\tau_l,\phi_l,\theta_l)$ for $l=1,...,L-1$ with $L=21$ and following 3GPP 3D-UMa NLOS model. The sum of gains is normalized to one.}
	\label{fig:gen_MPC}
\end{figure}
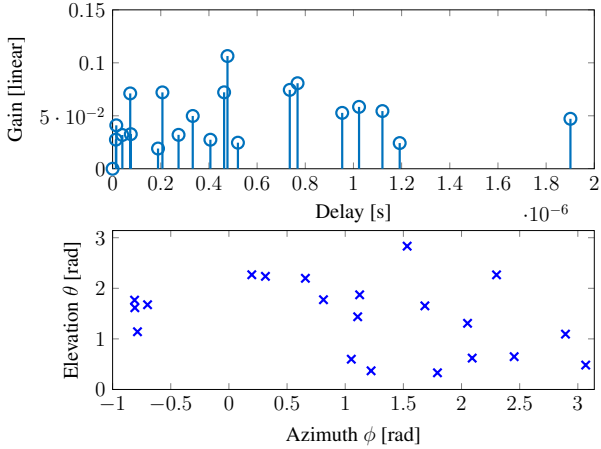
The MPC parameters were generated by the QuaDRiGa toolbox \cite{Jaeckel2014} according to the 3D-UMa NLOS model defined by 3GPP TR 36.873 v12.5.0 specifications \cite{3GPP_TR_36_873v12_5_0}. The same set of parameters was used for all simulations and is depicted in Fig.~\ref{fig:gen_MPC}. One can see that some rays are very closely spaced in delay and angle.

Three rectangular array geometries are considered: $M=8$ ($4\text{ Horiz.}\times 2\text{ Vert.}$), $M=32$ ($8\text{ Horiz.}\times 4\text{ Vert.}$) and $M=128$ ($16\text{ Horiz.}\times 8\text{ Vert.}$). Different values of the bandwidth, defined as $1/T$,\footnote{Note that, due to the roll-off factor, the exact bandwidth is actually $\frac{1+\beta}{T}$.} are considered as well. The performance in the figures is shown as a function of frequency normalized in the signal bandwidth $1/T$, as we expect form Corollary~\ref{corollary:well_separated_rays} that extrapolation scales accordingly. In the legend of the figures, the full LB refers to the LB of Theorem~\ref{theorem:LB} averaged over the receive antennas and the simplified LB refers to the expression of Corollary~\ref{corollary:well_separated_rays} if the rays are well separated. In the SISO case, the full and simplified LB curves refer to the corresponding expressions in Corollary~\ref{corollary:SISO_case}.

\subsection{SAGE performance versus theoretical LB}

\begin{figure}[!t]
	\centering
	\resizebox{0.45\textwidth}{!}{%
		\Large
		\input{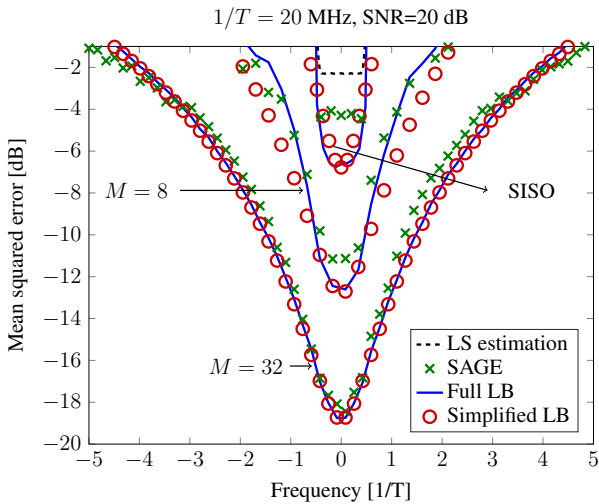}
	}
	\vspace{-1em}
	\caption{The SAGE algorithm can reach the performance of the full LB (Theor.~\ref{theorem:LB}). The simplified LB (Corol.~\ref{corollary:well_separated_rays}) gets closer to the full LB as the number of antenna increases meaning that $\mathbf{(As2)}$ is better validated: the rays are more separated in angle. We also notice a large SNR gain with respect to conventional LS.}
	\label{fig:accuracy_LB_reachable}
\end{figure}

Fig.~\ref{fig:accuracy_LB_reachable} shows the LBs of Theorem~\ref{theorem:LB}, Corollary~\ref{corollary:well_separated_rays} and Corollary~\ref{corollary:SISO_case}, the LS and SAGE estimation performance for the SISO, $M=8$ and $M=32$ cases. The first important point to notice is that SAGE can reach the performance of the full LB. This implies that the LB gives a good indicator of the achievable MSE. For the sake of clarity, we will omit SAGE performance in the next figures. Furthermore, the simplified LB gets closer to the full LB as the number of antenna increases meaning that $\mathbf{(As2)}$ is better verified: the rays are more separated in angle. The validity of $\mathbf{(As2)}$ will be studied in the following figures. In the SISO case,  $\mathbf{(As2')}$ is not at all valid and no extrapolation is possible. This will be studied in the following as well. As expected from the discussions of previous sections, the high-resolution channel estimators experience a large SNR gain with respect to conventional LS estimation and this gain scales with the number of antennas $M$.

\subsection{Influence of the bandwidth}

\begin{figure}[!t]
	\centering
	\resizebox{0.45\textwidth}{!}{%
		\Large
		\input{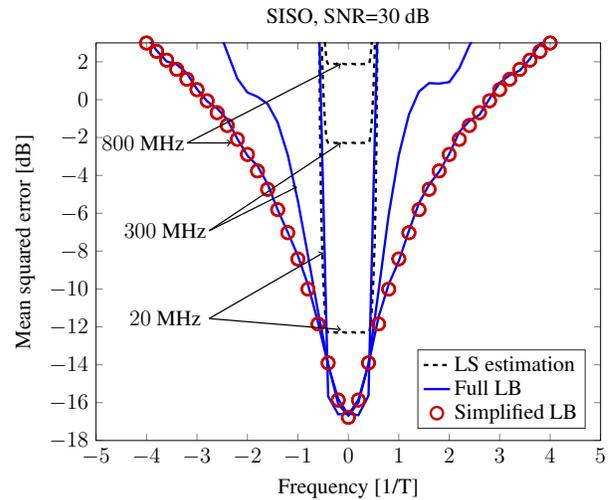}
	}
	\vspace{-1em}
	\caption{As the bandwidth increases, extrapolation becomes possible in SISO and the full LB converges to the simplified LB meaning that $\mathbf{(As2')}$ holds. A bandwidth of $1/T=800$ MHz is necessary for $\mathbf{(As2')}$ to be completely valid.}
	\label{fig:LB_fct_BW_SISO}
\end{figure}
\subsubsection{SISO}
As just explained, no extrapolation in SISO is possible in Fig.~\ref{fig:accuracy_LB_reachable} since the rays are too correlated or in other words, $\mathbf{(As2')}$ is not valid. Fig.~\ref{fig:LB_fct_BW_SISO} plots the evolution of the full LB as a function of the signal bandwidth for a SISO system. As the bandwidth of the system increases, the SISO system has a larger resolution in time and it can progressively resolve the different MPC. As the bandwidth increases, $\mathbf{(As2')}$ becomes more valid and the full LB converges to the simplified LB. The gap between the full and simplified LB can be seen as an indicator of the separability of the MPC.

As opposed to high-resolution channel estimation, increasing the bandwidth is detrimental to conventional LS estimation as the number of time domain coefficients to estimate becomes larger. Another way to view this is that the energy is more spread out in frequency and leads to a lower per-frequency bin SNR. We can conclude that the price to pay for channel extrapolation in SISO is to have a very large bandwidth at disposal and/or a channel that exhibits few well separated MPC. This observations tends to strongly limit the applicability of extrapolation for conventional SISO communication systems.

\begin{figure}[!t]
	\centering
	\resizebox{0.45\textwidth}{!}{%
		\Large
		\input{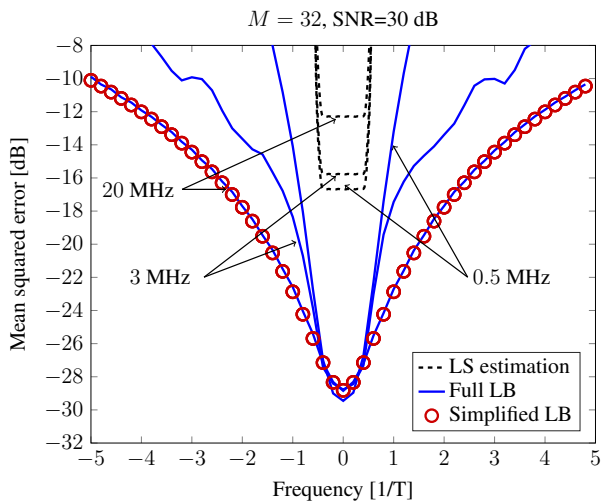}
	}
	\vspace{-1em}
	\caption{As opposed to SISO systems, for SIMO systems, the rays can be separated much more easily in the delay-angle domain, implying that $\mathbf{(As2)}$ becomes valid much faster. The maximal extrapolation gain is already attained for $1/T=20$ MHz while achieving an additional SNR gain.}
	\label{fig:LB_fct_BW}
\end{figure}
\begin{figure}[!t]
	\centering
	\resizebox{0.45\textwidth}{!}{%
		\Large
		\input{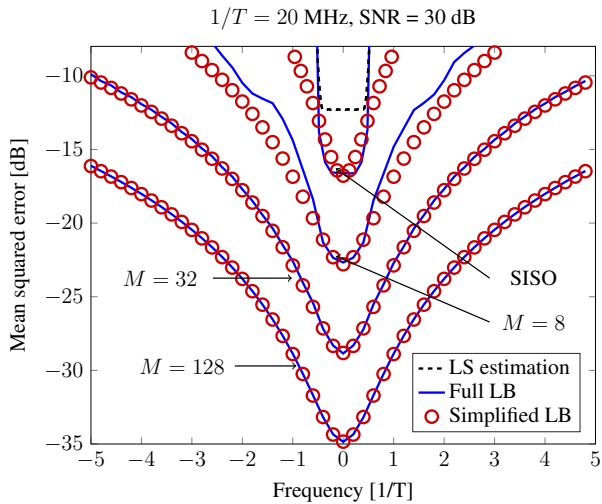}
	}
	\vspace{-1em}
	\caption{As the number of antenna elements increases, the rays can be more easily separated in the angle domain and the full LB converges to the simplified LB.}
	\label{fig:LB_fct_M}
\end{figure}

\subsubsection{SIMO}

The same type of remarks can be made for Fig.~\ref{fig:LB_fct_BW} which depicts the performance of a $M=32$ SIMO system for different bandwidths. As the bandwidth increases, assumption $\mathbf{(As2)}$ becomes more valid and the full LB converges to the simplified LB. The main difference with the SISO case is that the extrapolation becomes possible for much smaller bandwidth. This is explained by the fact that the rays can now be separated in the delay-angle domain and not only the delay domain. {In the end, we see that a 20 MHz SIMO system with $M=32$ antennas can reach the same extrapolation factor as a 800 MHz SISO system, with an additional SNR gain of 15 dB}.

\subsection{Influence of the number of antennas}

Fig.~\ref{fig:LB_fct_M} depicts the performance of the system for a fixed bandwidth of $1/T=20$ MHz as a function of the number of antennas. The same effect previously described occurs, \textit{i.e.}, as the number of antennas increases, the resolution in the angle domain increases and $\mathbf{(As2)}$ is more valid, the full LB converges to the simplified LB. These observations imply that the separability of the rays can be achieved by trading bandwidth against number of antennas.

\section{Conclusion} \label{section:conclusion}

This paper investigated the theoretical performance limits of channel extrapolation in frequency, seen as a potential solution to completely remove the pilot overhead for downlink channel estimation in FDD massive MIMO systems. We highlighted the advantages of basing the extrapolation on high-resolution channel estimation as compared to conventional estimators. A LB on the MSE of the extrapolated channel was proposed. By assuming that the rays are well separated, we were able to simplify the LB and directly identify the potential extrapolation range and SNR gain. It was shown that the SNR gain linearly scales with the number of receive antennas while the extrapolation performance penalty quadratically scales with the ratio of the frequency and the training bandwidth. From simulations using practical channel models, we saw that the derived LB can be reached with practical high resolution algorithms. Furthermore, the extrapolation range is very limited in SISO while much more promising results were obtained in SIMO as the paths can be separated in the delay-angle domain. Future works include validation of theoretical results by channel measurements and study of potential limiting factors such as channel modeling and calibration errors.

\section*{Acknowledgment}

The research reported herein was partly funded by the Belgian American Educational Foundation (B.A.E.F.).
\footnotesize
\bibliographystyle{IEEEtran}
\bibliography{IEEEabrv,IEEEreferences}

\end{document}